# ArEEG_Chars: Dataset for Envisioned Speech Recognition using EEG for Arabic Characters

Hazem Darwish, Abdalrahman Al Malah, Khloud Al Jallad, Nada Ghneim.

<u>201910306@aiu.edu.sy</u>, <u>201910477@aiu.edu.sy, k-aljallad@aiu.edu.sy</u>, <u>n-ghneim@aiu.edu.sy</u>

Department of Information and Communication Engineering,
Arab International University, Daraa, Syria.

## Abstract

Brain-computer interfaces is an important and hot research topic that revolutionize how people interact with the world, especially for individuals with neurological disorders. While extensive research has been done in EEG signals of English letters and words, a major limitation remains: the lack of publicly available EEG datasets for many non-English languages, such as Arabic. Although Arabic is one of the most spoken languages worldwide, to the best of our knowledge, there is no publicly available dataset for EEG signals of Arabic characters until now. To address this gap, we introduce ArEEG_Chars, a novel EEG dataset for Arabic 31 characters collected from 30 participants (21 males and 9 females), these records were collected using Epoc X 14 channels device for 10 seconds long for each char record. The number of recorded signals were 930 EEG recordings. To make the EEG signals suitable for analyzing, each recording has been split into multiple signals with a time duration of 250ms, respectively. Therefore, a total of 39857 recordings of EEG signals have been collected in this study. Moreover, ArEEG_Chars will be publicly available for researchers. We do hope that this dataset will fill an important gap in the research of Arabic EEG benefiting Arabic-speaking individuals with disabilities.

## 1. Introduction:

Human-Computer-Interface (HCI) is one of the most important fields in computer science that is concerned especially with the relationship between humans and computers.

Brain-Computer-Interface (BCI) is one of the HCI fields that has been a hot research topic in the last few decades. Especially researches that focus on using AI to classify EEG signals automatically where EEG signals are electrical activity in the brain measured by electroencephalography (EEG). Recent advancements in BCI show that brain signals are robust in decoding various mental tasks such as imagined speech, object understanding, etc. However, BCI in envisioned speech recognition using electroencephalogram (EEG) signals still have few researches.

Arabic language is a rich language with a long history and cultural significance. It is spoken by over 400 million people worldwide, and it is the official language in many countries. However, Arabic Language is still one of the most challenging low-resource languages.

In this paper, we present our method of creating ArEEG_Chars, an EEG dataset that contains signals of Arabic characters. We have reviewed the models used in the literature to classify the EEG signals, and the available datasets for English.

The main contribution of this paper is creating a dataset for EEG signals of all Arabic chars that will be publicly available for researchers[1]. To the best of our knowledge, ArEEG_Chars is the first Arabic EEG dataset.

The rest of the paper is organized as follows: Related works are presented in Section 2, Section 3 specifies the data collection method and the resulting dataset, and finally, section 4 gives the conclusion.

2. **Related Works:**

The first paper about EEG classification was published in 1924 by Hans Berger [1] where Berger described a method for classifying EEG signals into different categories based on their frequencies. This method was later used to develop EEG-based brain-computer interface (BCI).

We investigated the previous datasets and the existing models in detail. Some of the papers do not present their data collection process. We organized this section into "Benchmark datasets", and "Previous models" subsections.

**2.1. Benchmark Datasets**

There are several datasets for different languages that were collected using multiple EEG devices. The recognized units can be digits, chars, directions, audio, images, and words. In this paper, we will focus on Chars EEG datasets. Each study has a number of participants with different ages and various collecting methods. The number of participants varied from 3 to 29, the majority of the them were adults with good health conditions. As for the data collection method, there were many different scenarios of seeing the object for a duration of seconds then imagining the object, hearing the word they imagined in their minds, or even combining these methods. In some experiments, authors mentioned that the participants have to be free from any effects on their nervous system, such as coffee, alcohol, cigarettes, and so on.

In [2] Pradeep Kumar et al. developed envisioned speech recognition using EEG sensors, they collected data by EPOC+ device on 23 participants. A dataset of EEG signals has been recorded using 30 text and non-text class objects: letters, images and digits being imagined by multiple participants. As for the collection methodology, they recorded EEG signals by viewing an object on a screen then asked participant to imagine it with their eyes closed for 10 seconds. Before viewing the second object, the participants rested for 20 seconds. They collected 230 EEG samples per category for each participant. This produced 230*30 *23 samples.

---

[1] Almalah, Abdalrahman; Darwish, Hazem; Al Jallad, Khloud; Ghneim, Nada (2024), "ArEEG_Chars", Mendeley Data, V1, doi: 10.17632/f2bgr7tdtt.1, url: https://data.mendeley.com/datasets/f2bgr7tdtt/1

Nicolás Nieto in [3] entitled "Thinking out Loud", developed an open-access EEG-based BCI dataset for inner speech recognition using EEG 16, 17, 18, 19, 20, and 21 sensors on 10 participants (6 male, 4 female). The participant first heard one of 4 directions (up, down, left, and right), presented through a loudspeaker (average length: 800ms ± 20). A first cross was then displayed on the screen (1500ms after trial onset) for 1000ms, indicating that the participant had to imagine hearing the word. Finally, a second cross was displayed on the screen (3000ms after trial onset) for 1500ms, indicating that the participant had to repeat out loud the word. The words were chosen to maximize the variability of acoustic representations, semantic categories, and the number of syllables while minimizing the variability of acoustic duration.

In [4], researchers worked also on imagined speech classification for directions (up, down, left, right) using EEG and deep learning. They used Unicorn Hybrid Black + headset on 4 participants. The EEG dataset was acquired from eight EEG sensors and contained different frequency bands with different amplitude ranges. When the recording began, the question was announced after 10s to 12s as an audio pronounced by one of the other three participants. After 10 seconds, the participant started executing his response for 60 seconds by continuously imagining and saying the required command, and the recording was stopped after 10 seconds. In each recording, the participant responded by imagining saying the specified command, which was one of the four commands. Since there are four commands, the total recorded EEG dataset for all was 400 recordings.

In [5], seven participants were involved in the experiment, conducted in a dimly lit room where they minimized body and eye movements. EEG data was recorded using a 128-channel Geodesic Sensor Net with an ANT-128 amplifier at a 1024 Hz sampling rate. Their task was to imagine speaking one of two syllables, (/ba/ or /ku/), in one of three rhythms. The first rhythm has the time structure {| 1.5 | 1.5 |}, the second has the structure {| 0.75 | 0.375 | 0.375 | 0.75 | 0.375 | 0.375 |}, and the third has the structure {| 0.5 | 0.5 | 0.5 | 0.5 | 0.5 | 0.5 |}; the vertical lines "|" represent the expected times of imagined syllable production onset, while the numbers indicate the time intervals in seconds between imagined syllables. Each trial began with a 4.5-second cue period, where a syllable cue (/ba/ or /ku/) was presented for 0.5 seconds, followed by rhythmic cues. After a 1-second silence for baseline estimation, participants imagined speaking the syllable for 6 seconds. Each subject completed 720 trials across three rhythms and two syllables.

As for German language, Aurélie de Borman et al. in [6] entitled "Imagined speech event detection from electrocorticography and its transfer between speech modes and subjects," analyzed ECoG signals from 16 participants with subdural implants to detect imagined speech. Participants were instructed to perform three speech tasks: performed, perceived, and imagined speech. The Electrocorticography ECoG signals were recorded in seven frequency bands, from delta (0.5–4 Hz) to high-gamma (70–120 Hz). In each trial, participants were shown a short sentence, asked to memorize it, and then cued to perform one of the speech tasks. The short sentences imagined by the participants in the study were selected from the LIST database [7], which contains Dutch sentences. Sentences were presented randomly in one of two tasks vocalizing and listening.

Table 1 shows comparisons between EEG datasets.

| Type | Study | # Samples | # Participants | Age | Device | Frequencies | # Letter, Words & Images | Recording time |
|---|---|---|---|---|---|---|---|---|
| Digits & chars | [2] | 230 EEG sample for each per category for participant | 23 | 15-40 | 14 Channels EMOTIV EPOC | 2048Hz down-sampled 128Hz | 10 objects of each class (digits, alphabets, images) | 10 seconds |
| Directions | [3] | from 18 to 24 trials for each | 10 | 31 in average | 16, 17, 18 19. 20, and 21 sensors EEG headcap and the external electrodes | 1024Hz | 4 words | 5.8 seconds |
| Directions | [4] | / | 4 | 32 | Unicorn Hybrid Black+ | 250Hz | 4 words | 60 seconds |
| Vowels | [7] | 100 sample | 3 | 26-29 | 128 Channels Electrical Geodesics | 1024Hz | 2 vowels | 2 seconds |
| Vowels | [8] | / | 5 | 21-24 | 19 Channels | / | 5 vowels | / |
| Directions & Select | [9] | 33 epoch each subject | 27 | / | 14 Channels EMOTIV EPOC | 128Hz | 5 words | / |
| Directions & Select | [10] | 33 epoch each subject | 27 | / | 14 Channels EMOTIV EPOC | 128Hz | 5 words | / |
| Syllables | [5] | 120 trials | 7 | / | 128-channel Geodesic Sensor Net | 1024 Hz. | two syllables /ba/ and /ku/ | 10.5 seconds |
| Words | [6] | 60 trials per speech mode | 16 | / | Subdural electrodes, SD LTM 64 | 256 Hz | 20 Dutch words | 10.5 seconds |

| | | | | | Express amplifier | | | |
|---|---|---|---|---|---|---|---|---|

*Table 1 Comparison Between EEG Datasets*

## 2.2. Previous Models:

In the last few years, several researches were conducted to classify EEG signals automatically using machine learning and deep learning. In this section, we review these researches and make comparisons between the different approaches and results as shown in Figure 1. Figure 1 reviews the preprocessing techniques used to prepare the input signals, the methods used to extract the features from the pre-processed signal, and the models used in the classification step.

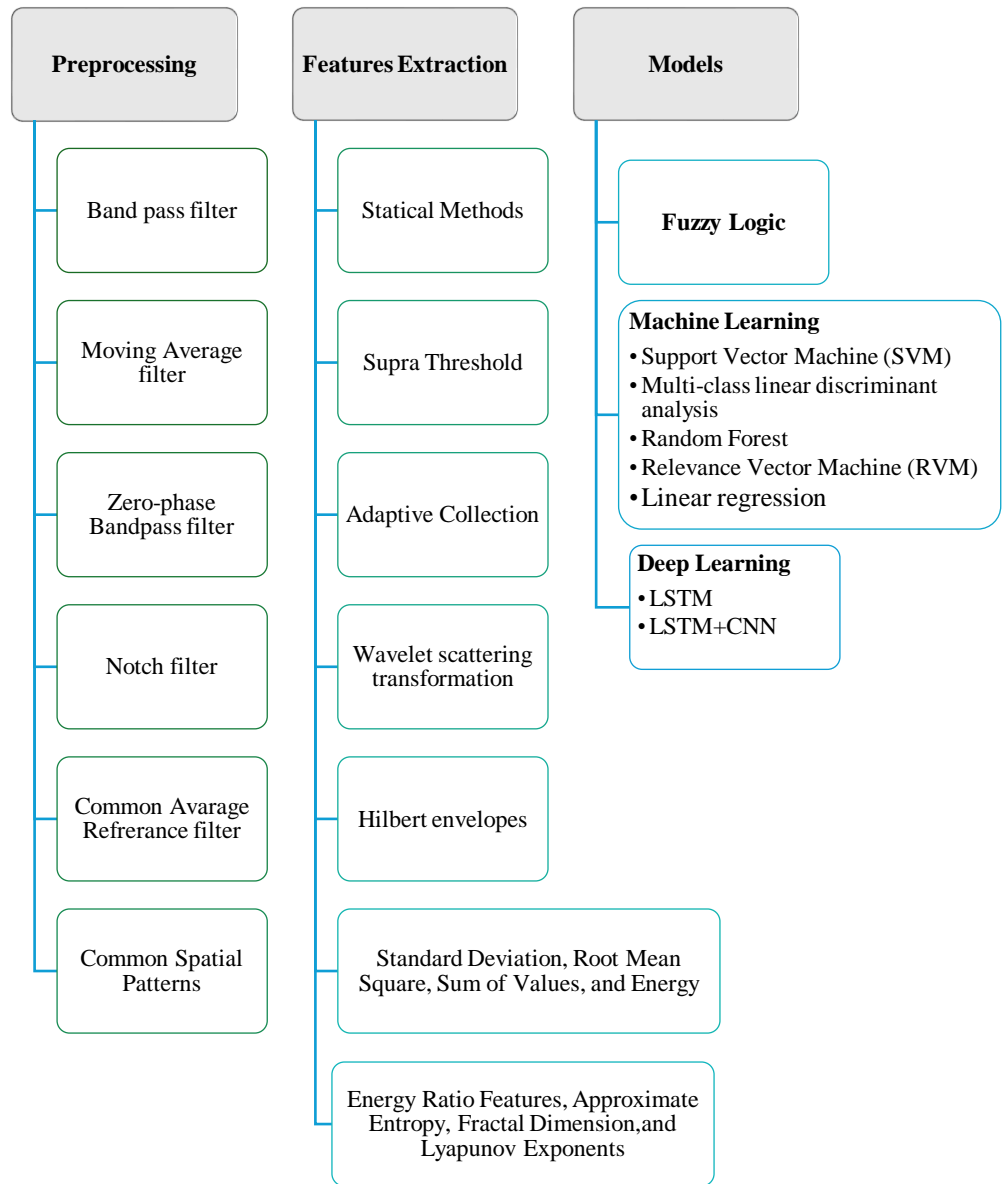

Figure 1 Our Review Summary of EEG Classification in previous Studies

DaSalla et al. [7] have proposed a BCI system to recognize English vowels using EEG signals of three subjects. The study has been conducted for three classes including two vowels and no-action state. A zero-phase bandpass filter has been applied within a frequency band of 1–45 Hz to remove lower frequencies and electronic noises. Common spatial patterns (CSP) have been applied to the EEG signals to generate new time series. An average accuracy of 71% has been recorded in all three classes using support vector machine (SVM) classifier.

The study in [5] employed several preprocessing techniques and models to achieve high classification accuracy for imagined speech rhythms of two syllables /ba/ and /ku/. Blind Source Separation (BSS) using the SOBI algorithm enhanced signal-to-noise ratio (SNR) by separating sources and reducing noise. Feature extraction involved transforming the Hilbert Spectrum (HS) into a binary representation to identify "spots of interest" based on frequency and time differences. Two models were used: the *SOBI-HHT algorithm*, which provided robust performance but had high computational complexity, and the *Spectrogram Matched Filter method*, which offered lower complexity but struggled with phase shifts,

In [9], Torres-García et al. have proposed a methodology for channel selection and classification of imagined speech using EEG signals focusing on the recognition of five Spanish words corresponding to the English words "up," "down," "left," "right" and "select". The selection of channels has been performed using fuzzy inference system (FIS) whose objectives was to minimize the number of channels and the error rate. Discrete Wavelet Transform (DWT) analysis has been performed on EEG data and five features including four statistical values and the relative wavelet energy (RWE) have been extracted for classification. The SVM has been used extensively for classification of EEG signals, and the evaluation was performed on a dataset of EEG signals from 27 subjects [12]. They reached 70.33% accuracy.

Pradeep Kumar et al. [2], used moving average filter, standard deviation root mean square, root mean square, sum of value and energy feature extractions to pre-process raw EEG data with Random Forest classifier and achieved an accuracy of 82.20% on a dataset that consists of three categories: images of words, letters and numbers.

The authors in [10] provide biometric security system. The Common Average Reference was applied to improve the signal-to-noise ratio. Two methods for feature extraction were used: The first, was based on discrete wavelet transform, and the second was based on statistical features directly from raw signals. The proposed methods were tested on a dataset of 27 Subjects who did 33 repetitions of 5 imagined words. As for classification, Random Forest classifier was used to classify inner speech. The accuracy obtained was different in each test case because it depends on the sensors number and feature extraction methods, and this accuracy ranged from 29% up to 96%.

Nicolás Nieto et al. [3] provided the scientific community with an open-access multiclass electroencephalography database of inner speech commands. Authors have used zero-phase bandpass filter using the corresponding Magnetoencephalography (MEG) and Electroencephalography (EEG) Neuroimaging (MNE) [11].

In [4], researchers used combination of Band pass Filter between 10 and 100 HZ and wavelet scattering transformation as feature extraction to preprocess the EEG data and achieved 92.74%, 92.50%, and 92.62% for precision, recall, and F1-score, respectively using LSTM model on a dataset of 4 inners speech words about directions.

In [6], Aurélie de Borman et al. have presented a methodology for detecting imagined speech events using ECoG signals, focusing on three speech modes: performed, perceived, and imagined speech. The signals were processed across seven frequency bands, from delta to high-gamma, and a linear regression model was employed for speech event detection. The Hilbert transform was used to extract signal envelopes, and their contributions were analyzed across different brain regions. Along with a single-electrode approach, they evaluated multi-electrode models and cross-validation using a leave-one-trial-out method. Additionally, they employed permutation tests to assess statistical significance, and transfer learning was explored across speech modes and between subjects. The study reached up to 75.03% accuracy for imagined speech detection in certain subjects.

As for Japanese language, in [8], authors have applied CSP and relevance vector machine (RVM) with Gaussian kernel for the classification of 5 imagined Japanese vowels. They compared three methods, support vector machine with Gaussian kernel (SVM-G), relevance vector machine with Gaussian kernel (RVM-G), and linear relevance vector machine (RVM-L). Results show that using RVM-G instead of SVM-G reduced the ratio of the number of efficient vectors to the number of training data from 97% to 55%.

Table 2 shows comparisons between EEG state of the art models.

| Ref | Preprocessing | | Model | Dataset | Accuracy |
|---|---|---|---|---|---|
| | Filter | Feature extraction | | | |
| [2] | Moving Average | Standard Deviation, Root Mean Square, Sum of Value and ENERGY | Random Forest | Consists of 3 categories (digits, alphabets, images) | 82.20% |
| [4] | Bandpass filter between 10 and 100 Hz | Wavelet scattering transformation | LSTM | Directions | 92.74%, 92.50%, and 92.62% for precision, recall, and F1score |
| [7] | A zero-phase bandpass filter Common Spatial Patterns | / | SVM | Two vowels | 71% |
| [8] | Special pattern filters (CSPS) filtering | Adaptive collection (AC) | RVM | 5 Japanese vowels | 79% |
| [9] | / | Energy ratio features, Approximate entropy, Fractal dimension, Lyapunov exponents. | FIS3×3 | 4 Directions and select | 70.33% |
| [10] | Common average reference | The first one is based on the discrete wavelet transform and the second one using statistical values Stander | Random Forest | | 29% to 96% |
| [6] | common average, band-pass filtering | Hilbert envelopes | Linear regression with leave-one-trial-out cross-validation | 20 Dutch words | 75.03% |

*Table 2 Comparison between models*

3. ArEEG_Chars Dataset

   3.1. Dataset Collection

   EEG recordings were collected using the Emotiv EPOC X. Thirty participants have been enrolled to collect data. Their ages ranged between 15 up to 62 years. All participants are educated and have been requested to remain calm during the whole process with clear thoughts. Moreover, all of them have been requested not to consume caffeine or alcohol and not to smoke before the recording process to avoid any effects of these substances on the nervous system.

   Collecting EEG signals took about an hour for each participant to record his thoughts about all Arabic chars. The participant set in a room with a comfortable chair and adjusted the headset on his/her head. A presentation that contained 31 Arabic letters was shown to participants, with a slide for each character. A pictorial representation of letters is shown in Figure 2. Letters slides in the presentation were randomly ordered to prevent the participant from thinking of the next letter. Each slide was shown to every participant for 10 seconds. Next, the participant was asked to envision the shown item for 10 seconds in the eyes-closed resting state. Between two successive recordings, a gap of 20 seconds has been introduced to clear the previous imaginary thoughts of the participant. To ensure that the participant imagined the letter as per the mentioned protocol, we asked the participant during the rest phase (at the beginning, middle, and end of the experiment) if he/she lost his/her focus while imagining the letter. The question helped us to keep the participants focused and having more responsibility about what they were imagining.

   Using this protocol, 930 EEG recordings have been collected. To analyze the EEG signals, each recording has been split into multiple signals with a time duration of 250ms, respectively. Therefore, a total of 39857 recordings of EEG signals have been analyzed in this study.

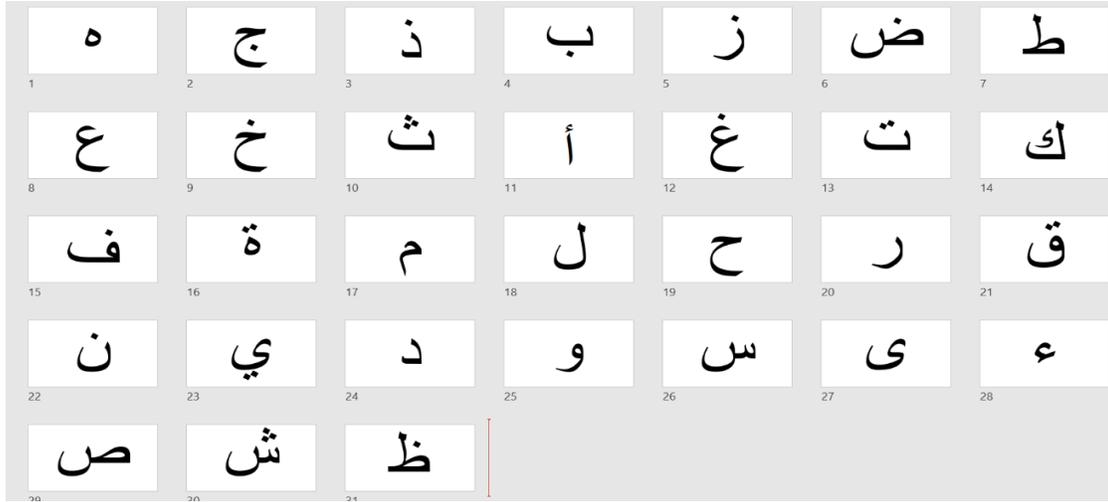

*Figure 2 Letters Used in the Study*

### 3.2. Statistics about our dataset

Thirty Arabic-native-speakers contributed in our experiments. Table 3 provides information about all participants.

| Participant No. | Age | Gender |
|---|---|---|
| **Par.1** | 16 | Male |
| **Par.2** | 16 | Male |
| **Par.3** | 23 | Male |
| **Par.4** | 21 | Male |
| **Par.5** | 22 | Female |
| **Par.6** | 22 | Male |
| **Par.7** | 24 | Male |
| **Par.8** | 23 | Male |
| **Par.9** | 24 | Male |
| **Par.10** | 22 | Male |
| **Par.11** | 28 | Female |
| **Par.12** | 28 | Female |
| **Par.13** | 24 | Female |
| **Par.14** | 29 | Male |
| **Par.15** | 24 | Male |
| **Par.16** | 23 | Male |
| **Par.17** | 37 | Male |
| **Par.18** | 31 | Female |
| **Par.19** | 40 | Female |
| **Par.20** | 38 | Female |
| **par.21** | 54 | Male |
| **Par.22** | 21 | Female |
| **Par.23** | 25 | Male |
| **Par.24** | 57 | Male |
| **Par.25** | 22 | Male |
| **Par.26** | 62 | Male |
| **Par.27** | 37 | Male |
| **Par.28** | 23 | Male |
| **Par.29** | 24 | Male |
| **Par.30** | 25 | Female |

*Table 3 Participants' information*

The majority of participants are male (70%). The average age is 26, with ages ranges between 16 and 64 as shown in Table 4. Moreover, the majority are high school students, as shown in Figure 3.

| Age Range | 1-19 | 20-39 | 30-39 | 40-49 | 50-59 | 60-69 |
|---|---|---|---|---|---|---|
| # Participants | 2 | 20 | 4 | 1 | 2 | 1 |

*Table 4 Participants Age Ranges*

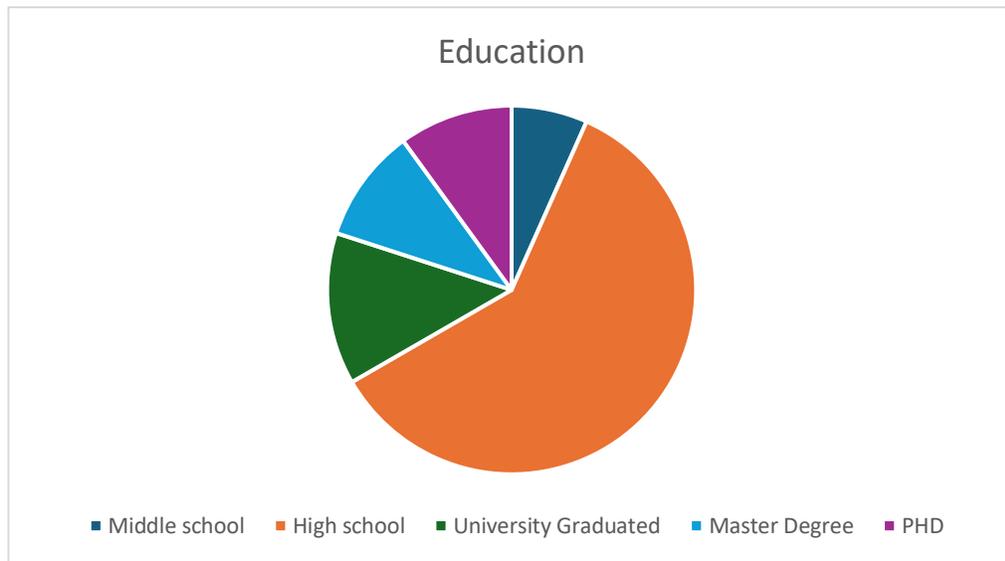

Figure 3 Participants Education Ratio

### 3.3. Headset:

In our framework, a wireless neuro-headset known as Emotiv EPOC X has been used for the acquisition of the envisioned brain signals. For recording brain signals, this device incorporates 14 channels namely AF3, F7, F3, FC5, T7, P7, O1, O2, P8, T8, FC6, F4, F8, AF4, which are placed over the scalp of the user according to the

International 10-20 system as shown in Figure 4, where two reference electrodes, i.e., CMS and DRL, are positioned above the ears.

We initially captured the brain waves at a frequency of 2048 Hz and later down-sampled them to 128 Hz per channel. The captured brain waves were sent to the computing device through Bluetooth technology.

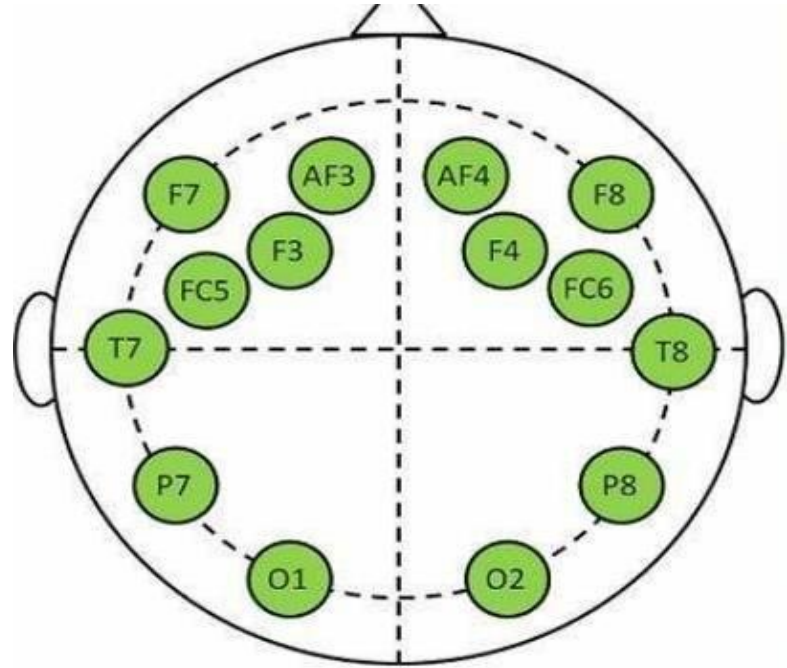

Figure 4 International 10-20 system

## 4. Conclusion

In conclusion, this work addressed a need in Brain-Computer Interface (BCI) research by creating ArEEG_Chars, the first publicly available dataset for electroencephalography (EEG) signals corresponding to Arabic characters. This meticulously constructed dataset, encompassing EEG data for all 31 Arabic characters, was collected from 30 participants from both genders using a 14-channel Epoc X device. ArEEG_Chars dataset serves as a foundation for future research, and to further propel BCI development for Arabic speakers, we are willing to create baseline models using deep learning to predict Arabic characters from the EEG signals. Furthermore, recognizing the importance of a richer data pool, we acknowledge the need for expanding ArEEG_Chars through the inclusion of additional participants. By making this dataset public, we aim to foster collaboration and accelerate advancements in BCI technology, particularly for Arabic native individuals with communication disabilities.